\documentstyle[12pt]{article}
\begin{document}
\newcommand{\st}{\stackrel}
\hsize=15.5cm
\textheight=24.5cm
\addtolength{\topmargin}{-90pt}
\large

\begin{center}

{\bf Solutions of Maxwell Equations for Hollow curved Wave
                 Conductor}\\

        V.Bashkov, A.Tchernomorov\\
     Kazan State University, Kazan, RUSSIA
\end{center}

\bigskip
{\bf Abstract}

                In the present paper the idea is proposed to solve
Maxwell equations for a curved hollow wave conductor by means
of effective Riemannian space,  in which the lines of  motion  of
fotons  are  isotropic  geodesies for a 4-dimensional space-time.
The algorythm of constructing such a metric and curvature tensor
components  are written down explicitly.The result is in accordance
with experiment.

 \bigskip
{\bf 1. Parallel translation of polarization vector in geometrical
optics of non-homologenous media.}

The discovery of law of parallel
translation for vectors $\st{\to}{e}=\frac{\st{\to}{E}}{\mid
\st{\to}{E} \mid}$ and $\st{\to}{h}=\frac{\st{\to}{E}}{\mid
\st{\to}{E} \mid}$
of  electric  and magnific fields in media with slowly changing
refraction factor is being rizen to the paper of S.M.Rytov [1] in
1938.
It was shown that for the light beam of nonflat curve form  there
is  a  rotation of vectors $\st{\to}{e}$ and $\st{\to}{h}$ in
respect to natural 3-facet,  created by tangent $\st{\to}{\tau}$,
normal $\st{\to}{n}$ and binormal $\st{\to}{b}$ vectors to the
curved beam . Setting :
\begin{equation}
\cases{
\st{\to}{e}=\st{\to}{n} \cos \varphi + \st{\to}{b}\sin \varphi,
     & $ $ \cr
\st{\to}{h}=-\st{\to}{n}\sin \varphi  + \st{\to}{b} \cos \varphi,
     & $ $}
\end{equation}
we find
\begin{equation}
     \frac {d\varphi}{ds}=\frac{1}{T},
\end{equation}
where $S -$ arch length,
         $T -$ tersion radius.Formula (2) is known as Rytov law.

        In 1941 the paper of V.V.Vladimirsky [2]  has  appeared
in  which on the basis of "Rytov Law" there was predicted a global
(topological) effect:  the angle of  rotation  of  polarized
light beam plane, trajectory of which in optically non-homogenous
media presents nonflat curve, equals integral of Gauss curvature
over  a region bounded by contour $C$, which is obtained by the top
of the vector $\st{\to}{\tau}$ on sphere of  unit  radius. The
angle $\theta$ equals space angle $\pi$  concluded inside of the
conus drawn by the
vector $\st{\to}{\tau}$. This angle is coincided with topological
Berry phase [3] for spiral fotons, experimentally  founded
in  recent works of Tomita and Chao [4] in spirally curved optical
wave conducters. In the case of a flat curve
$\theta= 0$ and after a beam gets initial direction the field
vectors take their previous position.
        In the first series of experiments the  spiral  was
uniformly winded on cylinder . In the second series of experiments
providing a fixed length P and radius r there were
used non-uniform spirals of different shapes, modeled
        on computer .  The space angle in this case is calculated
        by formula:
\begin{equation}
  \Omega (C)= \int^{2\pi}_0 \int^{\theta (\varphi)}_0 \sin \theta
d\theta d\varphi
 = \int^{2\pi}_0(1-\cos \theta)d\varphi
\end{equation}
Berry [3] was searching for solution
        of  classical  wave equations for a curved wave conductor
        by means of local coordinate system in  approximation  of
        idea of locally connected modes and weakly directing wave
        conductor [4]. It was produced not
taking into account radiation effects,  connection with reflected
modes and elastic-optical effects. It led in fact to the
approximation  of  geometricai  optics and could have been
analized in easier way,
taking into account that light trajectory represents a broken line,
corresponding to multiple reflections from walls [~5~],[~6~].
        In the  present  paper for interpritation of obzerved effect
we find the exact solutions of Maxwell equations for  hollow
curved  wave conductor in certain curved Riemannian space-time
different from flat Minkowsky space-time of special theory  of
relativity. The  choice  of 4-dimensional Riemannian space is such
that trajectories of light beams to be isotropic geodesics in
this  space, so the projection of this geodesics in
3-dimensional space were the lines corresponding to topology of  the
wave conductor.
        So consider a wave conductor the axis line  of  which  is
described by equation :
\begin{equation}
  \st{\to}{r}_0=\st{\to}{i}a \cos t + \st{\to}{j} a \sin t
 + \st{\to}{k} bt
\end{equation}
 in Cartesian coordinate system . It's easy to see that curvature
of  this  curve  is :
$$ K=\frac{1}{R}=\frac{a}{a^2+b^2}   $$
and a torsion : $\chi=\frac{1}{T}=\frac{b}{a^2+b^2}$.

Any point of wave conducter can be described by radius-vector
$\st{\to}{r}=\st{\to}{r}_0+\st{\to}{\rho}$, where
\begin{equation}
 \st{\to}{\rho}=\eta \st{\to}{n}+ \zeta \st{\to}{b}
\end{equation}
$\st{\to}{n}-$ normal vector to curve (5),
$\st{\to}{b}-$ binormal vector .
        The boundary of wave-conducter is described by equation
  $\rho=\rho_0.$  Let us introduce new variable
  $\xi \equiv s=\sqrt{a^2+b^2} t$ which  is  a  length  of  axis
line arch.Then  formula for any point inside the wave conductor
is :
\begin{equation}
\cases{
  x=(a-\eta)\cos{\frac{\xi}{\sqrt{a^2+b^2}}}-
  \zeta\frac{b}{\sqrt{a^2+b^2}}\sin{\frac{\xi}{\sqrt{a^2+b^2}}},
  & $ $ \cr
  y=(a-\eta)\sin{\frac{\xi}{\sqrt{a^2+b^2}}}+
 \zeta\frac{b}{\sqrt{a^2+b^2}}\cos{\frac{\xi}{\sqrt{a^2+b^2}}},
 & $ $ \cr
 z=\frac{b \xi- a \zeta}{\sqrt{a^2+b^2}}. & $ $ }
\end{equation}
Let's go from coordinates $( x,y,z )$ to
locally  curved  coordinates $( \xi,\eta, \zeta )$.
The metric tensor of 3-dimensional space in this coordinate
system being :
\begin{equation}
g_{\alpha \beta}=
\left( \begin{array}{ccc}
  1+\frac{K}{a}\eta^2+\chi^2\zeta^2-2K\eta &  \chi  \zeta  &  -  \chi
\zeta \\
  \chi \zeta & 1 & 0 \\
 -\chi \eta & 0 & 1
       \end{array} \right )
\end{equation}
 We suppose further that optical  beams $-$
lines of motion of fotons $-$ are curves,  which tangent
        vectors are orthogonal to planes $\xi=\xi_0$. It's
        dictated  by  our desire to get a non-dispersed beam at
        the exit of wave conductor. Congruation
of these lines is described by equations
\begin{equation}
\cases{
    \xi = \xi & $ $ \cr
    \eta = r \cos(\chi \xi+\varphi) & $ $ \cr
     \zeta = r\sin(\chi \xi + \varphi)}
\end{equation}
where $r$ and $\varphi$ -  some  parameters,
        characterizing the certain lines of congruation. Going to
        coordinates $(\xi,r,\varphi)$ we get for congruation lines
corresdonding to optical beams :
\begin{equation}
\cases{
  x=(a-r\cos(\chi \xi+\varphi))\cos{\frac{\xi}{\sqrt{a^2+b^2}}}-
  r\sin(\chi \xi +\varphi)\frac{b}{\sqrt{a^2+b^2}}\sin{\frac{\xi}
{\sqrt{a^2+b^2}}}, & $ $ \cr
  y=(a-r\cos(\chi \xi+\varphi))\sin{\frac{\xi}{\sqrt{a^2+b^2}}}+
 r\sin(\chi \xi +\varphi)\frac{b}{\sqrt{a^2+b^2}}\cos{\frac{\xi}
{\sqrt{a^2+b^2}}}, & $ $ \cr
 z=\frac{b \xi- a r \sin(\chi \xi +\varphi)}{\sqrt{a^2+b^2}}. & $ $ }
\end{equation}
Transition to 4-dimensional
space-time with  coordinates $(\xi, r, \varphi, t)$
        is accomplished
        by replacing $\st{0}{g}_{44}=1$ in Cartesian coordinate
        system in Minkowsky space on $g_{44}=-g_{11},$
so that in local coordinate system the metric tensor
of 4-dimensional space-time will be:
\begin{equation}
g_{ij}=
\left ( \begin{array}{cccc}
(1-rK\cos \phi)^2+\chi^2r^2  &  0  & -\chi r^2 & 0 \\
     0 & 1 & 0 & 0  \\
   -\chi r^2 & 0 & r^2 & 0 \\
    0 & 0 & 0 & -(1-rK \cos \phi)
       \end{array} \right )
\end{equation}
where $\phi \equiv \chi \xi+\varphi$.

        It's easy to show that congruation of isotropic geodesics
        of this space-time after projecting  gives the  lines
        of optical beams $-$ trajectories of fotons in
3-dimensional space, defined by formula (9). Non-vanishing
components  of  curvature  tensor of this Riemannian space-time
are:
\begin{equation}
\cases{
R_{1414}=K^2(1-rK\cos phi)+\chi^2K^2r^2 - \chi^2K^2\cos \phi & $ $ \cr
R_{1424}=\chi K \sin \phi & $ $ \cr
R_{1434}= -\chi K r(Kr-\cos \phi) & $ $}
\end{equation}

Ricci tensor components:
\begin{equation}
\cases{
R_{11}=K^2-\chi^2\frac{K^2(\cos \phi - Kr)}{(1-rK\cos \phi)^2} & $ $
\cr
R_{12}=\frac{\chi K \sin \phi}{(1-rK\cos \phi)^2} & $ $ \cr
R_{13}=\chi Kr\frac{(\cos \phi)-Kr}{(1-rK\cos \phi)^2} & $ $ \cr
R_{44}=-K^2-\chi^2\frac{K^2(\cos \phi- Kr)}{(1-rK\cos \phi)} & $ $}
\end{equation}

Scalar curvatur $R$ is:
\begin{equation}
R=\frac{2}{(1-rK\cos \phi)^2}{\Large [}{K^2+\chi^2
\frac{K^2(\cos \phi -Kr)}{(1-rK\cos \phi)^2}}{\Large ]}
\end{equation}

     And Einstein  tensor components:
\begin{equation}
\cases{
  G_{11}=-2\chi^2\frac{K^2(\cos \phi -Kr)}{(1-rK\cos \phi)^2}
-\frac{\chi^2r^2}{(1-rK\cos \phi)^2}{\Large [}{K^2+\chi^2
\frac{K^2(\cos \phi-Kr)}{(1-rK\cos \phi)^2}}{Large ]} & $ $ \cr
 G_{12}=\chi K \frac{\sin \phi}{(1-rK\cos \phi)^2} & $ $ \cr
 G_{13}=\chi \frac{K^2(\cos \phi-Kr)}{(1-rK\cos \phi)^2}+
 \frac{\chi r^2}{(1-rK\cos \phi)^2}{\Large [}{K^2+ \chi^2
\frac{K^2(\cos \phi-Kr)}{(1-rK\cos \phi)^2}}{\Large ]} & $ $ \cr
 G_{22}=-\frac{1}{(1-rK\cos \phi)^2}{\Large [}{K^2+\chi^2
\frac{K^2(\cos \phi-Kr)}{(1-rK\cos \phi)^2}}{\Large ]} & $ $ \cr
 G_{33}=-\frac{1}{(1-rK\cos \phi)^2}{\Large [}{K^2+\chi^2
\frac{K^2(\cos \phi-Kr)}{(1-rK\cos \phi)^2}}{\Large ]} & $ $}
\end{equation}

\bigskip
\bigskip
{ \bf 2.Solution of Maxwell equations.}

        In general  covariant  formulation  the Maxwell equations
are:
\begin{equation}
\cases{
    F_{[ij,k]}=0 & $ $ \cr
    F^{ij}_{,j}=0 & $ $ }
\end{equation}
where comma means covariant devivative,
and $F_{ij}-$ components of electromagnetic field tensor.
Writing down the system (15) in coordinates $(\xi, r, \varphi, t)$
in Riemannian space (10) it can be shown that it  has  a
partial solution corresponding to co-axial line with singularities
on axis line $r= 0$ of the type :
\begin{equation}
\cases{
     F_{12}=F_{24}=\frac{C_1}{r}e^{ik(\xi -t)}, & $ $ \cr
     F_{13}=F_{34}=C_2e^{ik(\xi -t)}, & $ $ \cr
     F_{14}=F_{23}=0. & $ $ }
\end{equation}

        For components of energy-momentum tensor of
electromagnetic field we find:
\begin{equation}
\cases{
     T_{11}=\frac{C^2_1+C^2_2}{r^2}, & $ $ \cr
     T_{14}=-\frac{C^2_1+C^2_2}{r^2}, & $ $ \cr
     T_{44}=\frac{C^2_1+C^2_2}{r^2}. & $ $ }
\end{equation}
\bigskip
{ \bf Conclusion.}

Returning  to  3-dimensional designations $\st{\to}{E}$
and $\st{\to}{H}$ by  means  of  the known formula it can
be shown from (16), that along fotons
lines (8) we found a vector of linear polarization
rotates on the angle $\chi\xi$,
what is well-coordinated with experimental data and caused by effect
of topological phase of Vladimirski-Berry.\\
\bigskip
{ \bf References.}\\

\noindent
1. Rytov S.M.,  Docl.  Academ.Nauc USSR, v.28,N 4 - 5, 1938, p.263.\\
2. Vladimirski V.V.,  Docl.  Academ. Nauc USSR, v.31,N 3, 1941,p.222\\
3. Berry M.V. Nature, 1987, v.326, N 6110, p.277\\
4. Tomito A.,Chaio R.V., Phis. Rev. Lett., 1986, v. 57, N8, p.937\\
5. Bialyniski-Birula I., Bialynski - Birula S., Phis.Rev.D., 1987,
   v.35, N8, p.2383\\
6. Vinitzki S.I., Derbov V.L., Dubrovin V.M., Marcovski B.L.,
Stepanovski Yu.P., Procedings  of  the work conference on elaboration
and construction of radiator and detector of gravitational waves,
Dubna, 1989, p.74.
\end{document}